# A Review of Power Aware Routing Protocols in Wireless Sensor Networks

*Sukhchandan Randhawa*
CSED
Thapar University
Patiala, India
Sukhchandan.95@gmail.com

*Anil KumarVerma*
CSED
Thapar University
Patiala, India
akverma@thapar.edu

***Abstract-*** WSNs are envisioned to consist of many small devices that can sense the environment and communicate the data as required. The most critical requirement for widespread sensor networks is power efficiency since battery replacement is not viable. Many protocols are proposed to minimize the power consumption by using complex algorithms. However, it is difficult to perform these complex methods since an individual sensor node in sensor networks does not have high computational capacity. On the other hand, many sensor nodes should transfer the data packet to the sink node that collects the required data. Therefore, the operations of the sensor nodes over the route are terminated. It is difficult to deliver the data packet to the sink node even if some sensor nodes are active.

In this paper, an introduction of WSNs is presented with a deep insight into the power-aware routing protocol for sensor networks. The protocols considered are – LEACH,VGA and PEGASIS. In addition, a comparison of these protocols is also presented.

***Keywords- Wireless Sensor networks, Power aware routing, Life time***

## I. INTRODUCTION

A WSN consists of a large number of low-cost, low–power sensor nodes that are deployed in a area of interest .Sensors have computation, communication, sensing capabilities. Sensor communicates via a short range radio signals and collaborate to accomplish the common tasks as shown in Fig. 1 [3] and having limited bandwidth, power, memory, processing resources and limited lifetime [2].

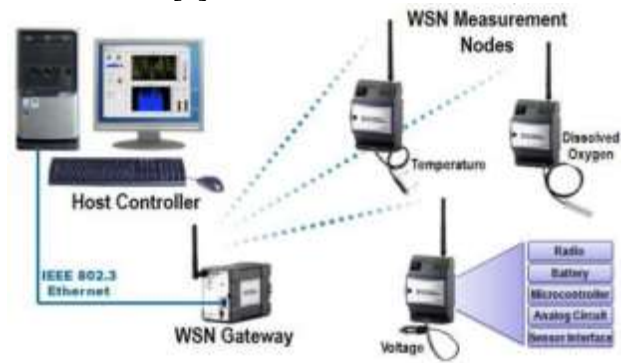

Figure 1: WSN[3]

## II. POWER SAVING MODES OF OPERATION

Sensor networks, must support power saving modes for the sensor node. For example means of power conservation is to turn the transceiver off when it is not required. Although this power saving method seemingly provides significant energy gains, an important point that must not be overlooked is that sensor nodes communicate using short data packets. The shorter the packets, the more the dominance of start up energy. In fact, if we blindly turn the radio off during each idling slot, over a period of time we might end up expending more energy than if the radio had been left on. As a result, operation in a power-saving mode is energy-efficient only if the time spent in that mode is greater than a certain threshold. There can be a number of such useful modes of operation for the wireless sensor node, depending on the number of states of the microprocessor, memory, A/D converter, and transceiver. Each of these modes can be characterized by its power consumption and latency overhead, which is the transition power to and from that mode. A dynamic power management scheme for wireless sensor networks is used where five power-saving modes are used as shown in table 1.

Table 1. Sleep States for the sensor nodes

| Sleep State | Processor | Memory | Radio | Sensor, Analog To Digital Convertor |
|---|---|---|---|---|
| S0 | Active | Active | Tx,Rx | On |
| S1 | Idle | Sleep | Rx | On |
| S2 | Sleep | Sleep | Rx | On |
| S3 | Sleep | Sleep | Off | On |
| S4 | Sleep | Sleep | Off | Off |

### III. POWER AWARE HIERARCHAL ROUTING PROTOCOL

Routing is one of the most critical tasks in any network and, therefore, a considerable amount of research has been conducted for traditional wired networks, cellular networks, ad-hoc networks with and without support for mobility and also wireless sensor networks. A number of routing algorithms have been proposed for wireless sensor networks including one-to-one, one-to-many, many-to-one and many-to-many routing tasks. The routing metric which is used to choose between alternative available paths in order to select the best one, where "best" is evaluated based on a predefined optimization goal.

Hierarchical routing performs energy-efficient routing in WSNs, and contributes to overall system scalability and lifetime. In a hierarchical architecture, sensors organize themselves into clusters and each cluster has a cluster head, i.e. sensor nodes form clusters where the low energy nodes are used to perform the sensing of the phenomenon. The less energy constrained nodes play the role of cluster-heads and process, aggregate and forward the information to a potential layer of clusters among themselves toward the base station. Now, there are three cluster based scheduling mechanisms.

#### A. LEACH Protocol

Heinemann, introduced a hierarchical clustering algorithm for sensor networks, called Low Energy Adaptive Cluster Hierarchy – protocol (LEACH) that utilizes randomized rotation of local cluster base stations (cluster-heads) to evenly distribute the energy load among the sensors in the network data aggregation reduces amount of information to be sent to base station; large reduction in energy dissipation as computation is much cheaper than communication can achieve as much as a factor of 8 in reduction in energy dissipation compared with conventional routing protocol.

In LEACH the operation is divided into rounds, during each round a different set of nodes are cluster-heads (CH) as shown in fig 2. Nodes that have been cluster heads cannot become cluster heads again for P rounds. Thereafter, each node has a 1/p probability of becoming a cluster head in each round. At the end of each round, each node that is not a cluster head selects the closest cluster head and joins that cluster to transmit data. The cluster heads aggregate and compress the data and forward it to the base station, thus it extends the lifetime of major nodes.

LEACH can be viewed as a hybrid approach using short and long range based data forwarding. The sensors within a cluster transmit their sensed data over short distances, whereas cluster heads communicate directly with sink. But this can be a problem so it is better to have multi-hop transmission instead of single hop transmission. In this algorithm, the energy consumption will distribute almost uniformly among all nodes and the non-head nodes are turning off as much as possible. LEACH assumes that all nodes are in wireless transmission range of the base station which is not the case in many sensor deployments. In each round, LEACH has cluster heads comprising 5% of total nodes. Fig. 2 shows the communications in LEACH protocol.

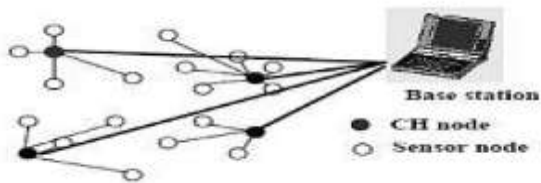

Figure 2 : LEACH [6]

### B. *PEGASIS Protocol[7]*

The protocol, called Power- Efficient Gathering in Sensor Information Systems (PEGASIS) is a near optimal chain-based protocol for extending the lifetime of network. The key idea in PEGASIS is to form a chain among the sensor nodes so that each node will receive from and transmit to a close neighbor. Gathered data moves from node to node, get fused, and eventually a designated node transmits to the BS. Nodes take turns transmitting to the BS so that the average energy spent by each node per round is reduced. It allows only cluster head to transmit their aggregated data to the sink in each round .A sensor has to transmit to its local neighbors in the data fusion phase instead of sending directly to its cluster head as in case of LEACH. It works by forming a chain first as shown in Fig. 3.

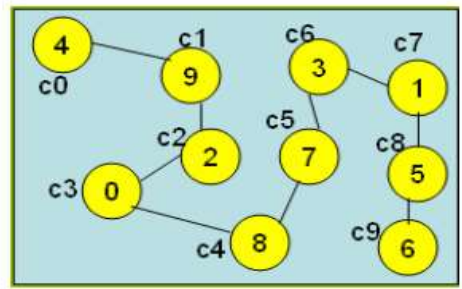

Figure 3: Chain formation in PEGASIS [8]

In PEGASIS, each node communicates only with the closest neighbor by adjusting its power signal to be only heard by this closest neighbor. Each Nodes uses signal strength to measure the distance to neighborhood nodes in order to locate the closest nodes. After chain Formation PEGASIS elects a leader from the chain in terms of residual energy every round to be the one who collects data from the neighbors to be transmitted to the base station. As a result, the average energy spent by each node per round is reduced.

Unlike LEACH, PEGASIS avoids cluster formation and uses only one node in a chain to transmit to the Base station instead of multiple nodes. This approach reduces the overhead and lowers the bandwidth requirements from the BS. Fig. 4 shows that only one cluster head leader node forward the data to the BS.

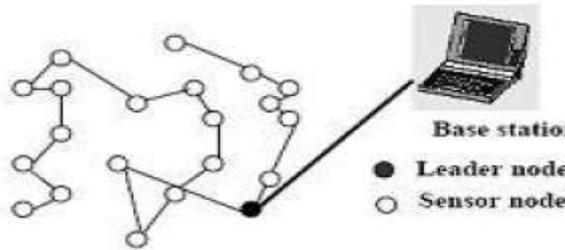

Figure 4: PEGASIS

A potential approach to reduce the delay required to deliver aggregated data to the sink is to use parallel data aggregation along the chain. For constructing the chain, we assume that all nodes have global knowledge of the network and employ the greedy algorithm. To construct the chain, we start with the furthest node from the BS. We begin with this node in order to make sure that nodes farther from the BS have close neighbors, as in the greedy algorithm the neighbor distances will increase gradually since nodes already on the chain cannot be revisited. It is assumed that nodes take turns in transmitting to the base station such that node i mod N, where N represents the total number of nodes, is responsible for transmitting the aggregate data to the base station in round i. Based on this assignment in fig 5, node 3, in position 3 in the chain, is the leader in round 3. All nodes in an even position must send their data to their neighbor to the right. At the next level, node 3 remains in an odd position. Consequently, all nodes in an even position aggregate their data and transmit them to their right neighbors. At the third level, node 3 is no longer in an odd position. Node 7, the only node beside node 3 to rise to this level, aggregates its data and sends them to node 3. Node 3, in turn, aggregates the data received with its own data and sends them to the base station.

PEGASIS improves on LEACH by saving energy in several stages. First, in the local gathering, the distances that most of the nodes transmit are much less compared to transmitting to a cluster-head in LEACH. Second, the amount of data for the leader to receive is at most two messages.

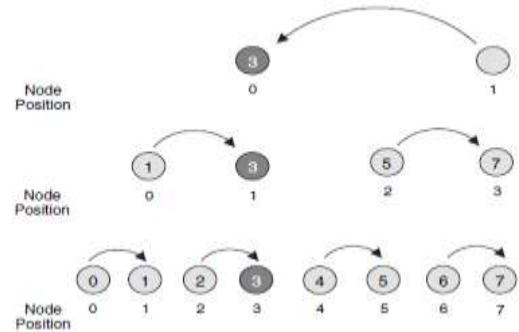

Figure 5: Chain based data gathering and aggregation scheme [1]

C. *VGA Protocol*

Virtual Grid Architecture (VGA) is an energy-efficient routing paradigm proposed in [8]. The protocol utilizes data aggregation and in-network processing to maximize the network lifetime.

Due to the node stationary and extremely low mobility in many applications in WSNs, a reasonable approach is to arrange nodes in a fixed topology.

A GPS-free approach is used to build clusters that are fixed, equal, adjacent, and non-overlapping with symmetric shapes. In [8], square clusters were used to obtain a fixed rectilinear virtual topology. Inside each zone, a node is optimally selected to act as CH.

Data aggregation is performed at two levels: local and then global. The set of CHs, also called Local Aggregators (LAs), perform local aggregation, while a subset of these LAs are used to perform global aggregation. However, the determination of an optimal selection of global aggregation points, called Master Aggregators (MAs), is NP-hard. Fig. 6 illustrates an example of fixed zoning and the resulting virtual grid architecture (VGA) used to perform two level data aggregation. Note that the location of the base station can be located at any arbitrary place.

## 1V COMPARISON AMONG PROTOCOLS

We have discussed three power aware protocols, but how can we choose that which one of them is best as per our requirement or whose performance is best. So in order to conclude the best one we need a comparison .The three power aware hierarchal protocols that we have discussed are compared below on the basis of some parameters like overhead, power usage, data aggregation, data delivery model, Query based, QoS etc as shown in the table below Table 2.

## SIMULATION [9][10]

Wireless sensor networks have the potential to become significant subsystems of engineering applications. Before relegating important and safety-critical tasks to such subsystems, it is necessary to understand the dynamic behavior of these subsystems in simulation environments. Different simulators such as ns2, GloMoSim, OPNET etc., are being used by researchers in order to evaluate the routing protocols. I have used ns2 for the evaluation of the proposed routing protocol as the same is an open source, freely available and the programming languages used are C++, Tcl and OTcl.

## CONCLUSION

WSNs are different kind of networks having their importance in certain areas such as Environment Monitoring, Military

Applications, and Health care applications, Industrial Process Control, Home Intelligence, Security and Surveillance etc.

A routing Protocol is used to decide on the best suitable route to be considered for sending data to the sink from a sensor node.

One of the major concern is to send this data on a route which consumes less power .The power is a scarce commodity in WSNs because when we deploy them in an hostile environment, it is not possible to give them power supply or to get them recharge. So there is need of key technologies required for low-energy distributed sensors.

These include power aware computation/communication component technology, low-energy signaling and networking, system partitioning considering computation and communication trade-offs, and a power aware software infrastructure and power aware Routing .An introduction about the different power aware routing Protocols such as LEACH, VGA, PEGASIS is presented and a comparison has also been carried out.

Table 2: Comparison among protocol

| Routing Protocols | Classification | Power Usage | Data aggregation | Data Delivery Model | Overhead | Scalability | QoS |
|---|---|---|---|---|---|---|---|
| LEACH | Hierarchical/ Node Centric | High | Yes | Cluster Head | High | Good | No |
| PEGASIS | Hierarchical | Max | No | Chain based | Low | Good | No |
| VGA | Hierarchical | Low | Yes | Good | High | Good | No |

REFERENCES

[1] D.Culler ,D. Estrin and M.Srivastava, "Overview of Sensor Networks", IEEE Computer Society , vol. 37 ,Issue No. 8, pp 41-49, August 2004 *(references)*

[2] I.F.Akyilidiz, W.Su, Y.Sankarsubramaniam and E.Cayirci. "A survey on Sensor Computer Networks', IEEE Communication magazine pp 102-114,August 2002.

[3] Zone.ni.com Assessed on 1.11.2011.

[4] W. R. Heinzelman, A. Chandrakasan, and H. Balakrishnan,"Energy-Efficient Communication Protocol for Wireless Microsensor Networks," *IEEE Proc. Hawaii Int'l. Conf. Sys. Sci.*, Jan. 2000, pp. 1–10.

[5] Laiali Almazaydeh, Eman Abdelfattah , Manal Al- Bzoor, and Amer Al-Rahayfeh Computer Science and Engineering Department "Performance Evaluation of Routing Protocols in WSN".

[6] L.B.Ruiz, J.M.Nogueria and A.A.F.Loureira, "Sensor network management",SMART DUST:*Sensor Network Applications ,Architecture and Design*, CRC Press ,Boca Raton ,FL,2006.

[7] Laiali Almazaydeh, Eman Abdelfattah , Manal Al- Bzoor, and Amer Al-Rahayfeh Computer Science and Engineering Department "Performance Evaluation of Routing Protocols in WSN".

[8] I. S. Jacobs and C. P. Bean, "Fine particles, thin films and exchange anisotropy," in Magnetism, vol. III, G. T. Rado and H. Suhl, Eds. New York: Academic, 1963, pp. 271–350.

[9] S. Lindsay and C. Raghavendra, "PEGASIS: Power-Efficient Gathering in Sensor Information Systems", *international Conf. on Communications, 2001. A new routing*

[10] Ian Downard Naval Research Laboratory Code 5523 "Simulating Sensor networks in NS2".